\documentclass[preprint,aps,showpacs]{revtex4}
\usepackage[dvips]{graphicx}

\newcommand{\blambda}{\mbox{\boldmath$\lambda$}}

\begin{document}

\title{On the Basis of Quantum Statistical Mechanics}

\author{A. Sugita
}
\email{sugita@a-phys.eng.osaka-cu.ac.jp}


\affiliation{
Department of Applied Physics,  Osaka City University,
      3-3-138 Sugimoto, Sumiyoshi-ku, Osaka 558-8585, Japan
}

\date{\today}

\begin{abstract}
 We propose a new approach to justify the use 
of the microcanonical ensemble for isolated macroscopic
quantum systems. Since there are huge number of independent
observables
in a macroscopic system, we cannot see all of them. Actually
what we observe can be written in a rather simple combination
of local observables.
Considering this limitation, we show that almost all states
in a energy shell are practically indistinguishable from one another, 
and hence from the microcanonical ensemble.
In particular, the expectation value of a macroscopic observable
is very close to its microcanonical
ensemble average for almost all states. 
\end{abstract}



\maketitle

\section{Introduction}
Justification of the principles of the equilibrium statistical mechanics
is still a controversial problem. For example, the widespread scenario
of obtaining the microcanonical
ensemble from the ergodicity
is not applicable to macroscopic systems,
because the "ergodic time" is too long \cite{Bricmont}.

In this paper we propose an alternative way to justify
the use of the microcanonical ensemble for macroscopic
quantum systems. The main point is that 
what we really observe can be written in
a rather simple combination of local observables.
For example, a macroscopic observable is additive
\cite{Shimizu_Miyadera},
and hence written as a sum of local observables
on a proper scale. 
Therefore we can ignore correlations among
macroscopic number of points.

Considering this limitation of observables, we show that almost all 
pure states in a energy shell look very similar.
Therefore we expect that the system go through only typical states
with high probability in the course of the time evolution
and thus thermal equilibrium is achieved. Note that strong
assumptions on the dynamics like ergodicity or mixing are
not necessary for a system to reach equilibrium in our picture.
It is possible even for an integrable system to reach equilibrium,
as has been pointed out in some earlier numerical works
\cite{Jensen_Shankar}.

It is known that a typical pure state is highly
(almost maximally) entangled in many senses \cite{Sugita_Shimizu}.
On the other hand, a thermal state is thought to have no or
little amount of entanglement. However, there is no contradiction
between the two viewpoints.
From our standpoint, a thermal state is a highly entangled
state, whose entanglement is too complicated to recognize.
In this sense, we can say that too much entanglement 
is no entanglement.

\section{Bloch vector}
To explain the idea more clearly, we introduce the Bloch vector as
a tool to represent quantum states. 
Let ${\cal H}_N$ be the Hilbert space of the system whose dimension is
$N$, and 
$V$ the set of traceless Hermitian operators (i.e., observables) on it.
We choose an orthonormal basis set $\left\{\hat{\lambda}_i | 1\le i \le N^2-1 \right\}$ 
of $V$.
Then the Bloch vector of a state is defined as
\begin{eqnarray}
\blambda \equiv \left(\langle\hat{\lambda}_1\rangle, 
\langle\hat{\lambda}_2\rangle,\dots,\langle\hat{\lambda}_{N^2-1}\rangle \right),
\end{eqnarray}
In this paper, we normalize the basis as
\begin{eqnarray}
{\rm Tr} \left(\hat{\lambda}_i \hat{\lambda}_j\right) = N\delta_{i,j},
\end{eqnarray}
so that the normalization condition for local observables 
does not depend on the
system size. The density matrix $\hat{\rho}$ and the Bloch vector
$\blambda$ are related as
\begin{eqnarray}
\hat{\rho} = \frac{1}{N}\left\{I + 
\sum_{i=1}^{N^2-1}\langle\hat{\lambda}_i\rangle \hat{\lambda}_i \right\}.
\end{eqnarray}
We use the $L_2$ norm for the Bloch vector
\begin{eqnarray}
\| \blambda \| \equiv \sqrt{\sum_i \langle\hat{\lambda}_i\rangle^2}
\end{eqnarray}
which is related to the Hilbert-Schmidt norm for the density matrix as
\begin{eqnarray}
\| \blambda_A - \blambda_B \| = 
\sqrt{N{\rm Tr}(\hat{\rho}_A - \hat{\rho}_B)^2}.
\end{eqnarray}
The set of the Bloch vectors is a subset of a ball with radius $\sqrt{N-1}$
in $R^{N^2-1}$. Pure states are on the surface of the ball, and mixed states
are in the interior. At the center, there is the completely
mixed state which is represented by the zero-vector $\blambda=0$.

\section{Projection to the space of relevant observables}

We consider a macroscopic system composed of $n$ sites, and
define a linear subspace of $V$ as
\begin{eqnarray}
V_m \equiv {\rm Span} 
\left\{\hat{a}_{\alpha_1}(l_1) \hat{a}_{\alpha_2}(l_2)\dots \hat{a}_{\alpha_k}(l_k)
| k\le m\right\}.
\label{basis}
\end{eqnarray}
Here, $m$ is a positive integer much smaller than 
$n$ but still a macroscopic number, 
and $\{\hat{a}_\alpha (l)\}$ is the basis of local operators at a site $l$.
We normalize the local observables as 
${\rm Tr}\left(\hat{a}_\alpha (l) \hat{a}_\beta (l)\right) 
= N_s \delta_{\alpha,\beta}$, where $N_s$ is the dimension
of the local Hilbert space. 
$V_m$ contains practically all observables of physical relevance. For example,
it contains all macroscopic observables and their
low-order moments.
We also define a projection operator for the Bloch vector $P_{V_m}$:
$P_{V_m}(\blambda)$ is the set of expectation values
of all basis elements of $V_m$.

We pick up a pure state in ${\cal H}_{[E,E+\Delta ]}$, 
which is a subspace of the Hilbert space spanned by eigenvectors
$|E_i\rangle$ with $E\le E_i \le E+\Delta$.
A state in ${\cal H}_{[E,E+\Delta ]}$ is written as
\begin{eqnarray}
|\psi\rangle = \sum_{i \in S} c_i| E_i\rangle,
\end{eqnarray}
where $S = \{i | E_i \in [E,E+\Delta]\}$.
If we choose the coefficients
$\{c_i\}$ randomly, 
the following inequality holds.
\begin{eqnarray}
\overline{\left\|P_{V_m}\left( \blambda\right) - P_{V_m}\left(\overline{\blambda}\right) \right\|^2}
\le \frac{N_s^m}{d + 1}\, {\rm dim}V_m.
\label{variance}
\end{eqnarray}
Here, $d = {\rm dim}{\cal H}_{[E,E+\Delta]}$ and the overline represents the
ensemble average over all states in ${\cal H}_{[E,E+\Delta]}$. 
With use of Chebyshev's inequality, one can also show
\begin{eqnarray}
{\rm Prob}\left( 
\left\| P_{V_m}\left( \blambda\right) - P_{V_m}\left(\overline{\blambda} \right) \right\|  
\ge k \right) 
\le 
\frac{N_s^m\, {\rm dim}V_m} {k^2(d + 1)}
\end{eqnarray} 
for any $k>0$.

As the system size $n$ increases,  the density of states increases exponentially
but ${\rm dim}V_m$ increases only in polynomial order $O(n^m)$.
Therefore the RHS of (\ref{variance}) is exponentially small for a macroscopic system,
which means that the portion of states 
practically distinguishable from the microcanonical ensemble is
exponentially small. 


To prove (\ref{variance}), we first prove the following important lemma
\begin{eqnarray}
\overline{\Delta \langle \hat{\lambda}\rangle^2}\le \frac{1}{d+1}|\hat{\lambda}|^2,
\label{lemma}
\end{eqnarray}
where $\Delta \langle\hat{\lambda}\rangle \equiv \langle\hat{\lambda}\rangle
- \overline{\langle\hat{\lambda}\rangle}$ and $|\hat{\lambda}|$ denotes 
the spectral norm of $\hat{\lambda}$.  
When we take the ensemble average over ${\cal H}_{[E,E+\Delta]}$, 
$\overline{|c_i|^2} = 1/d$, $\overline{|c_i|^4}=\frac{2}{d(d+1)}$,
and $\overline{|c_i|^2 |c_j|^2} = \frac{1}{d(d+1)}\;\;\; (i\ne j)$ 
\cite{Sugita_Shimizu, Ullah}. Up to the 4-th order, all other
combinations vanish identically. Therefore
\begin{eqnarray}
&&
\overline{\Delta \langle \hat{\lambda}\rangle^2}
\\
&=&
\overline{\langle\hat{\lambda}\rangle^2}
- \overline{\langle\hat{\lambda}\rangle}^2
\nonumber\\
&=&
\sum_{i,j,i',j'\in S}\lambda_{i,j}\lambda_{i',j'}
\overline{c_i^* c_j c_{i'}^* c_j}
- \left(\sum_{i,j\in S}\lambda_{i,j}\overline{c_i^* c_j}\right)^2
\nonumber\\
&=&
\frac{1}{d(d+1)}\sum_{i,j\in S}|\lambda_{i,j}|^2
- 
\frac{1}{d^2(d+1)}\left(\sum_{i\in S}\lambda_{i,i}\right)^2
\nonumber\\
&\le&
\frac{1}{d(d+1)}\sum_{i,j\in S}|\lambda_{i,j}|^2,
\label{intermediate}
\end{eqnarray}
where 
$\lambda_{i,j} = \langle E_i| \hat{\lambda} | E_j \rangle$.
Then
\begin{eqnarray}
\sum_{i,j\in S}|\lambda_{i,j}|^2
&\le&
\sum_{i\in S}\sum_{j}|\lambda_{i,j}|^2\\
&=&
\sum_{i\in S}\langle E_i|\hat{\lambda}^2|E_i\rangle\\
&=&
d\, \overline{\langle\hat{\lambda}^2\rangle}\\
&\le &
d\, |\hat{\lambda}|^2. 
\label{result}
\end{eqnarray}
Substituting (\ref{result}) to (\ref{intermediate}),
we obtain (\ref{lemma}). Then, since $|\hat{\lambda}^2|\le N_s^m$
for any basis element of (\ref{basis}), we obtain (\ref{variance}).

If we consider a small subsystem which consists of less than $m$ sites,
all observables in the subsystem is contained in $V_m$.
Since the microcanonical ensemble average of the density matrix
of a subsystem is the canonical ensemble in thermodynamically
normal systems, (\ref{variance}) gives a justification
of the canonical ensemble for a small subsystem.

\section{Time-dependent spin model}
Our idea can be illustrated clearly with a time-dependent
spin model. We consider the following Hamiltonian,
\begin{eqnarray}
H &=& 
J \sum_{l=1}^{N} \big\{
\sigma_{x}(l)\sigma_{x}(l+1)
+ \sigma_{z}(l)\sigma_{z}(l+1)
\nonumber \\
&&  \qquad
+ \sqrt{2} \cos \phi_{l} \, \sigma_{y}(l)\sigma_{y}(l+1)
\big\}
\nonumber \\
&&    
- h \sin (\omega t) \sum_{l=1}^{N} \{ \sin\theta_{l} \, \sigma_{x}(l)
+      \cos\theta_{l} \, \sigma_{z}(l) \},
\label{hamiltonian}
\end{eqnarray}
which is a time dependent version of the Hamiltonian used in \cite{Sugita_Shimizu}. Details of this Hamiltonian is not important here. The point
is that the time evolution is deterministic and there is no
constant of motion in this system.


If we start with a pure state, the state is always
pure and the length of the Bloch vector
is the maximum value $\sqrt{2^n-1}$. However,
the Bloch vector has so many components that
we never see all of them. When we see only 
small number of components, the Bloch vector
looks like the zero-vector with very high
probability because the average absolute value of each
component is very small.
Therefore the system looks like in the completely mixed
state, which is the equilibrium state expected from the principle
of equal a priori probabilities. Note, however, that
the equilibrium is obtained not
from any kind of averaging, but from the limitation
of observables.



\begin{figure}
\centering
\includegraphics[width=3.0in]{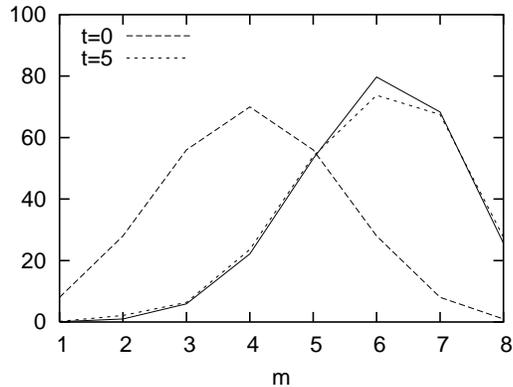}
\caption{Plot of $W_m$ for $n=8$ and
$J=h=\omega =1.0$.
The dashed and dotted lines
shows the results for $t=0$ and $t=5$
respectively.
The solid line shows the average over
all pure states.}
\label{fig}
\end{figure}

Fig. \ref{fig} is the plot of $m$-body part of 
the Bloch vector 
$W_m\equiv\left\|P_{V_m}(\blambda)-P_{V_{m-1}}(\blambda)\right\|^2$.
The initial state is a product state, which has relatively large
components for small $m$. In particular, the one-body part
takes the maximum value $W_1 = n$. After some time evolution
$W_m$ approaches quickly to the average value 
$W_m = \frac{3^m\, n!}{2^n\, m!(n-m)!}$,
which is very small if $m\ll n$. Therefore the system
looks like in the completely mixed state. 

We have also studied the transverse Ising model
\cite{Jensen_Shankar}
and cofirmed that the system approaches to equilibrium
even in the integrable case. Detailed analysis
of this model will be reported elsewhere.

\end{document}